\documentclass[nohyper]{JHEP3}

\usepackage{axodraw}
\usepackage{latexsym}
\usepackage{epsfig}

\def\beq{\begin{equation}}
\def\eeq{\end{equation}}
\def\beqa{\begin{eqnarray}}
\def\eeqa{\end{eqnarray}}
\def\eq#1{Eq.~(\ref{#1})}
\def\g{\gamma}

\def\ifm{\ifmmode}
\def\msb{\ifm \overline{\rm MS}\,\, \else $\overline{\rm MS}\,\, $\fi}
\def\msbns{\ifm \overline{\rm MS}\, \else $\overline{\rm MS}\, $\fi}
\newcommand{\bra}[1]{\langle{#1}|}
\newcommand{\ket}[1]{|{#1}\rangle}

\def\be{\begin{equation}}
\def\ee{\end{equation}}
\def\gtil{{\widetilde {G}}}

\def\summkp1{{\sum\limits_{m=1}^{k+1}}}
\def\summpp1{{\sum\limits_{m=1}^{p+1}}}
\def\summkm1{{\sum\limits_{m=1}^{k-1}}}
\def\summpm1{{\sum\limits_{m=1}^{p-1}}}

\def\a{\alpha}
\def\b{\beta}

\def\e{\epsilon}

\def\g{\gamma}

\def\m{\mu}

\def\p{\pi}

\title{Exponentiation of the Drell--Yan cross section
       near partonic threshold  in the $\mathrm{DIS}$ 
       and $\overline{\mathrm{MS}}$ schemes}

\author{Tim Oliver Eynck \\
        NIKHEF Theory Group, Kruislaan 409, 1098 SJ Amsterdam, 
        The Netherlands}
\author{Eric Laenen \\
        NIKHEF Theory Group, Kruislaan 409, 1098 SJ Amsterdam, 
        The Netherlands, and \\
        Institute for Theoretical Physics, Utrecht University, \\
        Leuvenlaan 4, 3584 CE Utrecht, The Netherlands}
\author{Lorenzo Magnea \\
        Dipartimento di Fisica Teorica, Universit{\`a} di Torino, and \\
        INFN, Sezione di Torino, Via P. Giuria 1, I-10125 Torino, Italy}

\abstract{It has been observed that in the DIS scheme the
refactorization of the Drell--Yan cross section leading to
exponentiation of threshold logarithms can also be used to organize a
class of constant terms, most of which arise from the ratio of the
timelike Sudakov form factor to its spacelike counterpart. We extend
this exponentiation to include {\it all} constant terms, and
demonstrate how a similar organization may be achieved in the \msb
scheme. We study the relevance of these exponentiations in a two--loop
analysis.}


\preprint{NIKHEF/2003--005 \\
          ITP--UU--03/15 \\
          DFTT--5/2003}

\begin{document}

\section{Introduction}
\label{intro}

The predictive power of QCD for high--energy hadronic scattering
observables rests upon our ability to compute the corresponding
partonic cross sections in perturbation theory. Factorization theorems
\cite{Collins:1989gx} assert that for many such cross sections mass
divergences may be subtracted, or mass--factorized, in a
process--independent way, with any additional finite subtraction
constants defining the mass--factorization scheme.

In many cases, the predictive power of such perturbative series is
imperiled by the systematic occurrence of finite but large terms at
higher orders. Resummation attempts to restore predictive power by
organizing classes of large terms to all orders, leaving a
perturbative series for the remainder with much better convergence
properties~\cite{Giele:2002hx}. For cross sections, the large terms
almost always take the form of logarithms of ratios of kinematical
scales. In particular, threshold
resummations~\cite{Sterman:1987aj,Catani:1989ne} organize logarithmic
enhancements singular at partonic threshold, resulting from imperfect
cancellations between real and purely virtual contributions to the
cross section.  As partonic threshold is approached, these
enhancements are parametrically guaranteed to increasingly dominate
the perturbative contributions to the cross section.  Often, however,
also constant terms, which do not depend on scale ratios vanishing at
threshold and which arise predominantly from purely virtual diagrams,
are numerically important in the cross section.  Some of these large
constants originate from the same infrared singularities that give
rise to the large logarithms, and consequently are resummable.

Threshold resummations can be derived for partonic cross sections
using the procedure of refactorization: the Mellin transform of the
cross section is expressed near threshold as a product of
well--defined functions, each organizing a class of infrared and
collinear enhancements. The refactorizations are valid up to
corrections which are nonsingular at threshold, and thus suppressed by
powers of the Mellin variable $N$ at large $N$. Terms independent of
$N$ can then be treated, in principle, by the same methods used to
resum terms enhanced by logarithms of $N$.  For the Drell--Yan cross
section and the deep--inelastic structure function $F_2$ such a
refactorization was achieved in Ref.~\cite{Sterman:1987aj}.  The
resulting resummation of constant terms established to all orders the
earlier observation~\cite{Parisi:1980xd} that, in the DIS--scheme
Drell--Yan cross section, the largest constants are related to the
ratio of the timelike Sudakov form factor to its spacelike
continuation.  By solving an appropriate evolution equation
\cite{Collins:1989bt}, an exponential representation for this form
factor, and thus also for the ratio, was derived~\cite{Magnea:1990zb},
generalizing the observation of~\cite{Parisi:1980xd} to a full
nonabelian exponentiation. Comparison with exact two--loop
results~\cite{Matsuura:1989sm} showed in that case that
$N$--independent contributions at two loops are indeed dominated by
the exponentiation of the one--loop result, combined with running
coupling effects~\cite{Magnea:1991qg,Magnea:1990db}.

In this paper we refine this analysis for the DIS scheme, and we
extend it to the \msb scheme. The challenge in the latter case lies in
the fact that the finite subtraction constants of this scheme are not
related to a physical scattering process involving the electromagnetic
quark coupling at lowest order. Therefore, ratios of form factors do
not naturally occur in the \msb scheme.  The practical prevalence and
relative simplicity of the \msb scheme would, however, make such an
organization desirable.  We will show that the refactorization
formalism of Ref.~\cite{Sterman:1987aj} leads to the exponentiation of
{\it all} $N$--independent contributions to the inclusive Drell--Yan
cross section, both in the \msb scheme and in the DIS scheme. As a
corollary, one may note that all constant terms in the \msbns--scheme
non--singlet deep--inelastic structure function $F_2$ have been
organized into an exponential form as well. Furthermore, it is
possible to organize the factorization procedure so that real and
virtual contributions are individually made finite; one can then
disentangle various sources of constants, such as $\pi^2$ terms
arising from unitarity cuts and similar terms arising from expansions
of phase--space related $\Gamma$ functions.

One might object that there is no kinematic limit in which
$N$--independent terms dominate parametrically, so that an
organization of such terms cannot be of much practical use. In
general, our view is that whenever all--order information is available
one should make use of it, at least to gauge the potential impact of
generic higher order corrections on the cross section at hand. One
should bear in mind that the pattern of exponentiation, even for
$N$--independent terms, is highly nontrivial, and includes all--order
information arising from renormalization group evolution and the
requirements of factorization; for example, a considerable fraction of
nonabelian effects arising in the Sudakov form factor at two loops can
be shown to follow from running coupling effects implemented on the
(essentially abelian) one--loop result. In the present case one could
be bolder and argue that, since all constants have been shown to
exponentiate, using the exponentiated expression should provide a
better approximation to the exact answer. This is in fact the case for
generic values of higher order perturbative coefficents, even for
asymptotic series such as those arising in QCD. It cannot, however, be
proven for any particular cross section, although it works in practice
for the cases that have been tested.  At the very least, differences
between results for the physical cross section with or without the
exponentiation of constant terms can provide nontrivial estimates of
errors due to (uncalculated) higher order corrections.  Notice in
passing that constant terms are not affected by the Landau pole and
thus factor out of the inverse Mellin transform needed to construct
the physical cross section.

This paper is organized as follows. In the next Section we review the
resummation of the quark Sudakov form factor and its embedding in the
DIS--scheme Drell--Yan cross section. We extend the exponentiation in
this scheme to include all constant terms. In Section 3 we derive our
result for the \msb cross section. In Section 4 we compare the results
of the exponentiations to finite order calculations, while Section 5
contains our conclusions. In an Appendix we present renormalization
group studies of certain auxiliary functions.

\section{Exponentiation in the DIS scheme} 
\label{DIS}

Consider the $N$--th moment of the partonic Drell--Yan cross section,
taken with respect to $z = Q^2/s$, with $Q$ the measured invariant
mass and $s$ the partonic invariant mass squared. Mass factorization
of this quantity, in the DIS scheme, is performed by simply dividing
its dimensionally regularized ($d = 4 - 2 \e$), unsubtracted version
by the square of the $N$--th moment (taken with respect to the
partonic Bjorken--$x$ variable) of the non--singlet partonic
deep--inelastic structure function $F_2$,
\beq 
\label{omdis}
\widehat{\omega}_{\mathrm{DIS}}(N) = \frac{\omega (N,
\e)}{\left[ F_2(N, \e) \right]^2} \, .  
\eeq 
While numerator and denominator are each infrared and collinear
divergent, their ratio is finite to all perturbative
orders~\cite{Collins:1989gx}, so the $\e$ dependence of the left hand
side can be neglected. Dependence on the hard scale $Q$ and on the
strong coupling $\a_s (\mu^2)$ will generally be understood.

Let us begin by identifying how the Sudakov form factor arises in
$\omega (N, \e)$, before mass factorization, following the reasoning
of Ref.~\cite{Sterman:1987aj}.  One observes that near threshold ({\it
i.e.}\ at large $N$) $\omega (N, \e)$ can be refactorized according to
\beqa
 \label{om}
 \omega (N,\e) = |H_{\mathrm{DY}}|^2 \, \psi(N, \e)^2 
 \, U(N) + {\cal O}(1/N) \, .
\eeqa
The $\psi(N, \e)$ and $U(N)$ functions contain the $N$ dependence
associated with initial state radiation at fixed energy and coherent
soft radiation, respectively.  They are well--defined as operator
matrix elements, as given in Ref.~\cite{Sterman:1987aj}, and are
calculable in perturbation theory. Both functions are
gauge--dependent, but their product in \eq{om} is not; implicitly, we
will be working in axial gauge.  Using gauge invariance and
renormalization group (RG) arguments, one can show that both the
parton distribution $\psi(N, \e)$ and the eikonal function $U(N)$ obey
evolution equations which can be solved near threshold in an
exponential form, up to corrections suppressed by powers of $N$. The
function $|H_{\mathrm{DY}}|^2$, collecting all hard--gluon
corrections, has no $N$ dependence and may be determined by matching
to exact calculations order by order. Divergences are only present in
the parton distribution function $\psi(N, \e)$.

To identify the Sudakov form factor, it is useful
\cite{Sterman:1987aj} to separate virtual ($V$) and real ($R$)
contributions to the resummed $\psi$ and $U$ functions, according to
\beqa
\label{revi}
 \psi(N, \e) & = & {\cal R}(\e) \: \psi_R(N, \e) 
 \nonumber \\
 U(N) & = & U_V(\e) \: U_R(N, \e)~, 
\eeqa
where ${\cal R}(\e)$ is the real part of the residue of the quark
two--point function in an axial gauge. Note that this further
factorization makes sense only because the functions $\psi$ and $U$
are known to exponentiate to the present accuracy. Using the analysis
of Section 8 of Ref.~\cite{Sterman:1987aj} we can now write
\beqa
\label{om2}
 \omega (N, \e) & = & |H_{\mathrm{DY}} \, {\cal R}
 (\e) \sqrt{U_V (\e)} |^2 \, \psi_R (N, \e)^2 \,
 U_R (N) + {\cal O}(1/N) \nonumber \\
 & = & |\Gamma(Q^2,\e)|^2 \, \psi_R(N, \e)^2 \,
 U_R(N, \e) + {\cal O}(1/N)~,
\eeqa
so that all virtual contributions are expressed in terms of the
electromagnetic quark form factor $\Gamma(Q^2, \e)$.  In fact, the
residue of the quark propagator coincides with the virtual jet
function summarizing virtual collinear contributions to the form
factor, while the square root of the virtual eikonal function
appearing in the cross section is responsible for the soft
enhancements of $\Gamma(Q^2, \e)$. We have thus indentified the
dimensionally regularized timelike Sudakov form factor in the
refactorized, unsubtracted Drell--Yan cross section. Near threshold,
the only remaining contributions to the cross section come from real
radiation, and are summarized by the real parts of the $\psi$ and $U$
functions. At this point one can already observe that
$\omega_{\mathrm{DY}}(N, \e)$ exponentiates up to corrections
suppressed by powers of $N$: the exponentiation of the form factor in
dimensional regularization was proven in Ref.~\cite{Magnea:1990zb},
while the exponentiation of $\psi_R$ and $U_R$ to this accuracy was
proven in Ref.~\cite{Sterman:1987aj}. Specifically, the real part of
the fixed--energy parton density $\psi_R(N, \e)$ can be written as
\beq
\label{epsir}
 \psi_R(N, \e) = \exp \left\{ \int_0^1 dz ~\frac{z^{N - 1}}{1 - z} 
 \int_z^1 \frac{dy}{1 - y} ~\kappa_\psi \left(\overline{\a} 
 \left((1 - y)^2 Q^2 \right), \e \right) \right\}\,.
\eeq
Similarly
\beq
\label{eur}
 U_R(N, \e) = \exp\left\{- \int_0^1 dz ~\frac{z^{N - 1}}{1 - z} \,
 g_U \left( \overline{\a} \left((1 - z)^2 Q^2\right), \e 
 \right) \right\}\,.
\eeq
Note that in writing \eq{epsir} and \eq{eur} one makes use of the fact
that both functions are renormalization--group invariant, since real
emission diagrams do not have in this case ultraviolet
divergences. For the function $\psi_R$ this is a consequence of the
fact that it fixes the energy of the final state, so that transverse
momentum is also limited; for the function $U_R$ it is a consequence
of the structure of nonabelian exponentiation, as discussed in
Ref.~\cite{Sterman:1987aj}. The functions $\kappa_\psi$ and $g_U$ are
both finite at one loop in the limit $\e \to 0$: all IR and collinear
singularities are generated by the integrations over the scale of the
$d$--dimensional running coupling, which at one loop is given by
\beq
\label{alphadim}
 \overline{\a} \left(\frac{\xi^2}{\mu^2}, \a_s(\mu^2), \e
 \right) = \a_s(\mu^2) \: \left[ \left(\frac{\xi^2}{\mu^2} \right)^{2
 \e} - \frac{1}{\e} \left\{ 1 - \left(\frac{\xi^2}{\mu^2}
 \right)^{2 \e} \right\} \: \frac{b_0}{4\pi} \a_s(\mu^2) 
 \right]^{-1} \, ,
\eeq
and will often be abbreviated by $\overline{\a} (\xi^2)$, as in
Eqs.~(\ref{epsir}) and (\ref{eur}).

A similar refactorization can be performed on the deep inelastic
structure function $F_2$ \cite{Sterman:1987aj}. One finds
\beq
\label{F2}
 F_2 (N, \e) = |H_{\mathrm{DIS}}|^2 \, \chi(N, \e) 
 \, V(N) \, J(N) + {\cal O}(1/N) \, .
\eeq
Here the parton distribution $\chi(N, \e)$ has the same soft and
collinear singularities as the one adopted for the Drell--Yan process,
but, according to the general strategy of Ref.~\cite{Sterman:1987aj},
it fixes a different component of the incoming parton momentum (the
plus--component) and is computed in a different axial gauge. It
exponentiates in a form similar to \eq{epsir}, but with a different
function $\kappa_\chi$ appearing in the exponent. Note that
$\kappa_\psi$ and $\kappa_\chi$ must (and do) differ at one loop only
by terms of order $\e^2$, so that the ratio $\psi_R/\chi_R$ may remain
finite.  The soft function $V(N)$ summarizes the effects of coherent
soft radiation in the DIS process, and exponentiates in a form similar
to \eq{eur}. Finally, $J(N)$ contains the effects of final state
collinear radiation emitted by the struck parton, and separately
exponentiates in a somewhat more elaborate form which will not be
needed here.  Separating real and virtual contributions as above we
find
\beqa
\label{F2fac}
 F_2 (N, \e) & = & |H_{\mathrm{DIS}}|^2 \, |{\cal R}(\e) 
 V_V(\e)| \, \chi_R(N, \e) \,
 V_R(N, \e) \, J(N) + {\cal O}(1/N) \nonumber \\
 & = & |H_{\mathrm{DIS}}| |\Gamma(- Q^2, \e)| \, 
 \chi_R(N,\e) \, \sqrt{V_V(\e)} \, V_R(N, \e) 
 \, J(N) + {\cal O}(1/N)~.
\eeqa
To this extent virtual contributions to $F_2$ have been organized
already in \cite{Sterman:1987aj}.  We now see that \textit{all}
virtual contributions can be organized in terms of $\Gamma(- Q^2,
\e)$, by observing that the purely virtual part of the light--cone
distribution $\chi$ is identical to the purely virtual part of the
outgoing jet $J$. Both consist essentially of the full two--point
function for a lightlike fermion. Note that both virtual jets in
\eq{F2fac} are computed with the same gauge choice. Gathering all
virtual parts, one finds then
\beq
 \label{F2g}
 F_2 (N, \e)  =  |\Gamma(- Q^2, \e)|^2 \, \chi_R(N, 
 \e) \, V_R(N, \e) \, J_R(N,\e) + {\cal O}(1/N) \,.
\eeq
Again, exploiting the results of
Refs.~\cite{Sterman:1987aj,Magnea:1990zb} this form of the
refactorization is sufficient to prove the exponentiation of the full
cross section up to corrections suppressed by powers of $N$.  In fact
$F_2$ now involves, to this accuracy, only the form factor, and a
product of real functions which have been shown to exponentiate by
using their respective evolution equations.

This result can be further verified in the following way.  A
comparison of Eqs.~(\ref{F2fac}) and (\ref{F2g}) implies that
$H_{\mathrm{DIS}}$ itself acquires an exponential form. In fact, an
analysis along the lines of \cite{Collins:1989bt} reveals that it may
be expressed as
\beq
 \label{hexp}
 H_{\mathrm{DIS}}(Q^2) = Z_H(\a_s,\e)\, \frac{\Gamma(-Q^2,
 \e)}{S^{(0)}(\e)\,\, G^{(0)}_2(Q^2,\e)}~,
\eeq
where $S^{(0)} = S \, Z_S$ and $G^{(0)}_2 = G_2 Z_q$ are the
unrenormalized, dimensionally regularized virtual soft function and
virtual quark jet function appearing in the factorization of the form
factor, and where
\beq
 \label{ZH}
 Z_H = Z_S \, Z_q^2
\eeq
is the UV counterterm function for $H_{\mathrm{DIS}}$. The Sudakov
form factor $\Gamma$ does not require a separate $Z$ factor to cancel
QCD UV divergences, by virtue of electromagnetic current conservation.
Now, each factor in \eq{hexp} has an exponential form.  For the
virtual soft function this was shown by Gatheral, Frenkel and Taylor
in \cite{Gatheral:1983cz,Frenkel:1984pz}. The unrenormalized virtual
jet function $G^{(0)}_2(Q^2,\e)$ obeys an evolution equation of the
same form as the one used for the full form
factor~\cite{Collins:1989bt}, which can be explicitly solved in
dimensional regularization by the same methods, using as initial
condition the fact that all radiative corrections vanish at $Q^2 = 0$.
$G^{(0)}_2(Q^2,\e)$ must then exponentiate by itself.  Finally, any
$Z$ factor arising in multiplicative renormalization may be
represented in a minimal scheme in terms of the associated anomalous
dimension $\gamma = (1/2) d (\ln Z)/d\ln\mu$ as
\beq
 \label{Zi}
 Z_i = \exp\left\{\int_0^{Q^2}\frac{d\xi^2}{\xi^2} 
 \gamma_i\left( \overline{\a} \left(\xi^2 \right) \right) \right\}~,
\eeq
where again UV poles are generated by integration over the scale of
the $d$--dimensional coupling.

Turning to the evaluation of \eq{omdis}, we observe that it requires
the ratio of \eq{om2} and the square of either \eq{F2fac} or
(\ref{F2g}).  In practice, the expression (\ref{F2fac}) is more
convenient than \eq{F2g}, because the resulting form (\ref{findis}) for
$\widehat{\omega}_{\mathrm{DIS}} (N)$ is a product of finite
functions. Had one used instead the result in \eq{F2g}, the resulting
expression for $\widehat{\omega}_{\mathrm{DIS}} (N)$ would have
involved cancelling divergences between the real and virtual
parts. Using then \eq{F2fac}, and the additional information that $U_V
(\e) = V_V (\e)$, both being given by pure counterterms to the same
eikonal vertex, we can write
\beq
\widehat{\omega}_{\mathrm{DIS}} (N) = \frac{1}{|H_{DIS}|^2}
\left|\frac{\Gamma(Q^2, \e)}{\Gamma(-Q^2, \e)} \right|^2
\left(\frac{\psi_R(N, \e)}{\chi_R(N, \e)} \right)^2 
\frac{U(N)}{V^2(N)} \frac{1}{J^2(N)}~.
\label{sumdis}
\eeq
The exponentiation and RG running of the various factors of
\eq{sumdis} are described in detail in \cite{Sterman:1987aj}, with the
exception of the ratio $\psi_R/\chi_R$, which there was exponentiated
according to \eq{epsir}, but evaluated only at leading order.  Running
coupling effects on this ratio are briefly discussed in an Appendix:
they generate a contribution at NNL log level at two loops, as well as
further $N$--independent terms. Furthermore, we are now in a position
to exponentiate also the one--loop contribution to the matching
function, setting $H_{DIS} = \exp \left(- \a_s C_F/\pi \right)$.
Gathering all factors, and formulating the answer according to
standard notation~\cite{Catani:1989ne}, our result for the hard
partonic Drell--Yan cross section in the DIS scheme takes the form
\beqa
\label{findis}
 \widehat{\omega}_{\mathrm{DIS}} (N) & = & 
 \left|\frac{\Gamma(Q^2, \e)}{\Gamma(-Q^2, \e)} \right|^2
 \exp \Big[ F_{DIS} (\a_s) \Big]
 \nonumber \\ 
 & \times & \exp \Bigg[ \int_0^1 \! dz \,
 \frac{z^{N - 1} - 1}{1 - z} \Bigg\{ 2 \, \int_{(1 - z)Q^2}^{(1 - 
 z)^2 Q^2} \frac{d \xi^2}{\xi^2} \, A \left( \a_s( \xi^2) 
 \right) \nonumber \\
 & - & 2 B \left(\a_s \left((1 - z) Q^2 \right) \right) + 
 D \left( \a_s \left((1 - z)^2 Q^2 \right) \right) 
 \Bigg\} \Bigg] + {\cal O}(1/N) \, .
\eeqa
\eq{findis} resums all terms in the perturbative expansion which
contain enhancements of the form $\a_s^n \log^k N$, with $k \leq 2 n$,
provided the perturbative expansions of the functions $A$, $B$ and $D$
are known to the desired order in the strong coupling.  The
perturbative coefficients of the functions $A$, $B$, $D$ are in fact
all known up to two
loops~\cite{Contopanagos:1997nh,Vogt:2000ci,Moch:2002sn}\footnote{Concerning
the three-loop coefficient $A^{(3)}$, the $n_f$--dependent part is
known exactly \cite{Moch:2002sn,Berger:2002sv}, while good numerical
estimates exist for the $n_f$--independent
term~\cite{vanNeerven:2000wp}.}.  For example, expanding the functions
involved as
\beq
\label{pert}
 f(\a_s) \: = \: \sum_{k = 1}^{\infty} \, f^{(k)} 
 \left( \frac{\a_s}{\pi} \right)^k \: , 
\eeq
one needs 
\beqa
 A^{(1)} = C_F  & , & \qquad 
 A^{(2)} = \frac{1}{2} \left[C_A C_F \left(\frac{67}{18} - \zeta(2) \right) - 
 n_f C_F \left(\frac{5}{9} \right)\right]  \nonumber \\
 B^{(1)} = - \frac{3}{4} C_F  & , & \qquad D^{(1)} =  0~.
\label{coeff}
\eeqa
for resummation to next--to--leading logarithmic accuracy. 

\eq{findis} also exponentiates $N$--independent terms, which have
three sources: they come from unitarity cuts, as in the analytic
continuation of the form factor; or from phase space integrations,
since for example the parton distributions $\psi$ and $\chi$ differ
slightly in their phase space measure; finally they arise from the
Mellin transformation in the exponent, which generates not only
logarithms of $N$, but also contributions proportional to $\gamma_E$
and to $\zeta(n)$. Let us examine in more detail the first two classes
of exponentiated constants.

As far as the ratio of form factors is concerned, one may use the
result of Ref.~\cite{Magnea:1990zb} showing that the absolute value of
the ratio is finite to all orders and exponentiates. To illustrate
these results, note that the (timelike) Sudakov form factor
$\Gamma(Q^2, \e)$ for the electromagnetic coupling of a massless quark
of charge $e_q$ is defined via
\beq 
 \Gamma_\mu (p_1, p_2; \mu^2, \e) \equiv \bra{0}
 J_\mu (0) \ket{{p_1,p_2}} \label{def} = - {\rm i} e\, e_q ~\overline{v}(p_2) 
 \gamma_\mu u(p_1) ~\Gamma \left( Q^2, \e \right)~, \nonumber
\label{ffdef}
\eeq
with $Q^2 = (p_1 + p_2)^2$.  Based on
Refs.~\cite{Sen:1981sd,Mueller:1979ih,Collins:1980ih}, it was shown in
Ref.~\cite{Magnea:1990zb} that the dimensionally regularized Sudakov
form factor may be written as an exponential of integrals over
functions only of $\a_s$,
\beq
\label{mygam}
 \Gamma(Q^2, \e) = \exp \Bigg\{
 \frac{1}{2} \int_0^{Q^2} \frac{d\xi^2}{\xi^2} \Bigg[
 K \left(\a_s, \e \right) +
 G \left(\overline{\a} \left(\xi^2 \right), \e \right) + 
 \frac{1}{2} \int_{\xi^2}^{\mu^2} \frac{d \lambda^2}{\lambda^2}
 \gamma_K \left( \overline{\a} \left(\lambda^2\right) \right) 
 \Bigg] \Bigg\} .
\eeq
The function $K$ in \eq{mygam} is defined to consist of counterterms
only, while $G$ is finite for $\e \to 0$. Double logarithms of the
hard scale $Q$ arise from the double integral over the anomalous
dimension $\gamma_K (\a_s)$. Note that within the framework of
dimensional regularization all integrals in \eq{mygam} can be
explicitly performed using the $d$--dimensional running
coupling~\cite{Magnea:2000ss}, so that the form factor can be
expressed in terms of RG--invariant analytic functions of $\a_s$ and
$\e$ to any order.

Using \eq{mygam} one can derive a particularly simple expression for
the absolute value of the ratio appearing in \eq{findis}. One finds
\beq
 \left|\frac{\Gamma(Q^2, \e)}{\Gamma(-Q^2, \e)} \right| =
 \exp \left\{ \left| \frac{\rm i}{2} \int_0^\pi \left[ G \left(
 \overline{\a} \left( {\rm e}^{{\rm i} \theta} Q^2 \right),
 \e \right) -
 \frac{\rm i}{2} \int_0^\theta d \phi\, \gamma_K  \left( \overline{\a} 
 \left( {\rm e}^{{\rm i} \phi} Q^2 \right) \right) \right] \right| 
 \right\}~.
\label{genrat}
\eeq
Performing the scale integrals, at the two--loop level this yields
\beqa
\label{rattwo}
 \left| \frac{\Gamma(Q^2, \e)}{\Gamma(-Q^2, \e)} \right|^2 & = & 
 1 \: + \: \frac{\a_s(Q)}{\pi} \: \frac{3 \zeta(2) \gamma^{(1)}_K}{2} 
 \\ & + & \left(\frac{\a_s(Q)}{\pi}\right)^2 \: \left[\frac{9}{8} 
 \zeta^2(2) \left(\gamma^{(1)}_K\right)^2 + \frac{3}{4} \zeta(2) 
 b_0 G^{(1)}(0) + \frac{3}{2} \zeta(2) \gamma^{(2)}_K \right]~, \nonumber
\eeqa
where
\beqa
\label{pgg}
 \gamma^{(1)}_K & = & 2 \, C_F~, \nonumber \\
 G^{(1)} (\e) & = & C_F \left(\frac{3}{2} - \frac{\e}{2} 
 \left[\zeta(2) - 8 \right] + \e^2 \left[8 - \frac{3}{4} \zeta(2) -
 \frac{7}{3} \zeta(3) \right] + {\cal O}(\e^3) \right)~, \nonumber \\
 \gamma^{(2)}_K & = & C_A C_F \left(\frac{67}{18} - \zeta(2) \right) - 
 n_f C_F \left(\frac{5}{9} \right) \, = \, 2 A^{(2)},
\eeqa
while $b_0 = (11 C_A - 2 n_f)/3$. The anomalous dimension $\gamma_K$
is the ``cusp'' anomalous dimension of a Wilson line in the \msb
renormalization
scheme~\cite{Kodaira:1982nh,Korchemsky:1993xv,Korchemsky:1992zp}.

\eq{rattwo} illustrates the potential relevance of exponentiation of
$N$--independent terms: first of all, the two--loop contribution is
numerically dominated by one--loop effects, both through
exponentiation and the running of the coupling (the first two terms of
the two--loop coefficient numerically make up roughly three quarters
of the total); furthermore, for this particular ratio, genuine
two--loop effects are given only in terms of $\gamma^{(2)}_K$, thus
they are UV--dominated and much simpler to calculate than the full
form factor.

Finally, the function $F_{DIS} (\a_s)$ collects constant terms arising
from phase space integrations in the various functions involved in the
factorization, as well as from the exponentiation of the matching
function $H_{DIS}$. One finds at one loop
\beq
F_{DIS}^{(1)} = C_F \left( \frac{1}{2} + \zeta_2 \right) \quad ,
\label{fdis}
\eeq
while at two loops some terms can be predicted by taking into account
the running of the coupling, which yields
\beq
F_{DIS}^{(2)} = - \frac{3}{16} C_F b_0 \left(4 + \zeta(2) - 2 
\zeta(3) \right) + \delta F_{DIS}^{(2)}~.
\label{fdis2}
\eeq
These two--loop contributions should not be taken too literally since,
as indicated in \eq{fdis2}, there is at this level an uncalculated
contribution $\delta F_{DIS}^{(2)}$ arising from a pure two--loop
calculation, which could easily overwhelm the effects which have been
included. Further discussion of the impact of these two--loop effects
is given in Section 4.

\section{Exponentiation in the $\overline{\rm MS}\,\, $ scheme}
\label{sec:msb-scheme}

As remarked in the introduction, one should not expect that the
constants associated with the Sudakov form factor in the
\msbns--scheme Drell--Yan cross section can be organized as in the
previous Section, in terms of a simple ratio of the timelike to
spacelike versions of the same form factor.  The reason, of course, is
that the \msb quark distribution function is not directly related to a
physical process involving quark electromagnetic scattering at lowest
order. We will show that it is nevertheless possible to organize these
constants in a closely related manner.

Mass factorization of the Drell--Yan cross section in the \msb scheme
is straightforward in moment space: one simply divides $\omega(N, \e)$
by $\phi_{\msb}^2(N, \e)$, the square of the \msb quark distribution,
defined by
\beq
\label{phims}
 \phi_{\msb}(N, \e) = \exp\Bigg[
 \int_0^{Q^2}  \frac{d \xi^2}{\xi^2} \, \Bigg\{
 \int_0^1 \! dz \, \frac{z^{N - 1} - 1}{1 - z}
 A \left( \overline{\a} \left(\xi^2 \right) \right)
 + B_\delta \left(\overline{\a} \left(\xi^2 \right) 
 \right) \Bigg\} \Bigg] + {\cal O}(1/N)\, ,
\eeq
with $A$ the same function appearing in \eq{findis}, while $B_\delta$
is the virtual part of the non--singlet quark--quark splitting
function ($B_\delta (\a) = B(\a)$ at one loop).  As appropriate for an
\msb parton density, one can easily verify that $\phi_{\msb}(N, \e)$
in \eq{phims} is a series of pure counterterms.  $Q$ is the
factorization scale, which for simplicity throughout this paper we set
equal to the Drell--Yan invariant mass~\footnote{It is straightforward
to repeat the analysis below keeping these scales different.}. We will
now factor this density into virtual and real parts
\beq
\label{phivr}
 \phi_{\msb}(N, \e) =  \phi_V(\e) \: 
 \phi_R(N, \e)\,,
\eeq
in such a way that in the (finite) ratio 
\beq
\label{omms}
 \widehat{\omega}_{\msb}(N) \equiv \frac{\omega(N,
 \e)}{\phi_{\msb}(N, \e)^2} = 
 \left(\frac{|\Gamma(Q^2, \e)|^2}{\phi_V(\e)^2} \right) \:
 \left(\frac{\psi_R(N, \e)^2 \, U_R(N, \e)}{\phi_R(N,
 \e)^2} \right) + {\cal O}(1/N) 
\eeq
the ratios of virtual functions and of real functions, displayed in
the large brackets, are separately finite. To be precise, the
factorization in \eq{phivr} is uniquely defined by the following
criteria: first, the ratio of virtual functions must be finite;
second, as we are factorizing a series of pure counterterms, we would
like also $\phi_V(\e)$ to consist only of poles. The real part
$\phi_R(N, \e)$ is then defined by \eq{phivr}.  Note that $\phi_V(\e)$
defined in this way is process--dependent, in contrast to
$\phi_{\msb}(N, \e)$; note also that, while $\phi_{\msb}$ has only
simple poles of collinear origin, the real and virtual contributions
will have cancelling double poles. We will now analyze separately the
real and virtual contributions to the cross section.

\subsection{Cancellation of virtual poles}
\label{sec:virt-canc}

The timelike Sudakov form factor has imaginary parts, which are the
source of the largest contributions to the ratio in \eq{genrat}, while
the \msb distribution is real. We can thus simplify our analysis by
writing
\beq
\label{splr}
 \frac{|\Gamma(Q^2, \e)|^2}{\phi_V(\e)^2} 
 = \left|\frac{\Gamma(Q^2, \e)}{\Gamma(-Q^2, \e)} \right|^2
 \left(\frac{\Gamma(-Q^2, \e)}{\phi_V(\e)} \right)^2~. 
\eeq
The benefit of this lies in the fact that the first factor on the
right hand side is finite, and already explicitly resummed in
\eq{genrat}. The second factor, on the other hand, is purely
real. Inspired by the explicit expression for the form factor,
\eq{mygam}, we try the ansatz
\beq
\label{phiv}
 \phi_{V} (\e) = \exp \Bigg\{
 \frac{1}{2} \int_0^{Q^2} \frac{d\xi^2}{\xi^2} \Bigg[
 K \left(\a_s, \e \right) +
 \gtil \left(\overline{\a} \left(\xi^2\right) \right) +
 \frac{1}{2} \int_{\xi^2}^{\mu^2} \frac{d \lambda^2}{\lambda^2}
 \gamma_K \left( \overline{\a} \left(\lambda^2 \right) 
 \right) \Bigg] \Bigg\}~,
\eeq
which has the same structure as the Sudakov form factor in \eq{mygam},
with the difference that $\gtil$ has no order $\e$ terms.

We will now show, to all orders in the strong coupling, that the
perturbative coefficients of the function $\gtil$ can be chosen so as
to render \eq{splr} finite, and we will provide for them an explicit
construction. Since the first ratio on the right hand side of
\eq{splr} is finite, it is sufficient to prove the cancellation of
poles for the second ratio, $\Gamma(-Q^2, \e)/\phi_V(\e)$. First of
all, one observes that all divergences arising from the terms $K$ and
$\g_K$ manifestly cancel between $\Gamma(-Q^2, \e)$ and $\phi_V(\e)$,
since they come from limits of integration independent of the
particular energy scale considered. Hence divergences could only be
generated by the $G$ terms. To show that those terms can be made
finite as well, we begin by writing
\beq
 G \left(\a_s, \e \right) = \sum_{n = 1}^\infty \sum_{m = 0}^\infty
 G_m^{\left( n \right)} \e^m \left({{\a_s}\over{\p}}\right)^n~,
\label{gexp}
\eeq
and by noting that the integration over the energy scale can be
rewritten as one over the running coupling making use of
\beq
 {{d\m^2}\over{\m^2}} = 2 ~{{d\,\overline{\a}}\over{\b \left(\overline{\a},
 \e \right)}} = - \frac{1}{\e} \frac{d \overline{\a}}{\overline{\a}} 
 \frac{1}{1 + \frac{1}{4\e}
 \sum\limits_{m = 1}^\infty b_{m - 1} \left(\frac{\overline{\a}}{\p} 
 \right)^m}~.
\label{ovpol}
\eeq
To proceed, let us define the truncated perturbative expansions for
the various functions involved by
\beqa
 \widehat{\beta}_{(M)}(\a_s, \e) & \equiv & 1 + \frac{1}{4 \e} 
 \sum_{n = 1}^M b_{n - 1} \left( \frac{\a_s}{\pi} \right)^n~, 
 \nonumber \\ G_{(M)} \left(\a_s, \e \right) & \equiv & \sum_{n = 1}^M 
 \sum_{m = 0}^\infty G_m^{\left( n \right)} \e^m \left({{\a_s}\over{\p}}
 \right)^n~, \\
 \widetilde{G}_{(M)} \left(\a_s \right) & \equiv & \sum_{n = 1}^M
 \widetilde{G}^{(n)} \left({{\a_s}\over{\p}}\right)^n~. \nonumber
\label{trunca}
\eeqa
Finally, define
\beq
 R_{(M)}^{(p)} (\a_s, \e) \equiv \frac{G_{(M)} \left(\a_s, \e \right) -
 \widetilde{G}_{(M)} \left(\a_s \right)}{\widehat{\beta}_{(M - 
 1)}(\a_s, \e)}~,
\label{rest}
\eeq
where the index $(p)$ denotes the fact that the perturbative expansion
for the ratio $R$ is truncated at order $\a_s^p$.

Writing down the integrands of both the numerator and the denominator
of the second ratio in \eq{splr} and keeping in mind the overall
factor of $1/\e$ from \eq{ovpol}, one can easily formulate our goal in
terms of $R$. We need to show that $\gtil$ can be chosen so that
\beq
 R_{(M)}^{(M)} (\a_s, \e) = {\cal O} (\e) \qquad , \qquad \forall M~.
\label{thes}
\eeq
The proof proceeds by induction. First of all, one sees immediately
that it is possible to get $R_{(1)}^{(1)} (\a_s, \e) = {\cal O} (\e)$,
simply by choosing $\gtil^{(1)} = G_0^{(1)}$. Next we assume that
\eq{thes} holds for some fixed order $M$ in perturbation theory for
chosen values of $\gtil^{(n)}$, up to $n = M$, and show that
$\gtil^{(M + 1)}$ may be chosen so that \eq{thes} holds also to order
$M + 1$.

To isolate the terms of order $\a_s^{M + 1}$ in $R_{(M+1)}^{(M+1)}$,
we use the identity
\beq
 \frac{1}{\widehat{\beta}_{(M)}(\a_s, \e)} = 
 \frac{1}{\widehat{\beta}_{(M - 1)}(\a_s, \e)} - 
 \frac{\widehat{\beta}_{(M)}(\a_s, \e) - \widehat{\beta}_{(M - 1)}( 
 \a_s, \e)}{\widehat{\beta}_{(M)}(\a_s, \e) \widehat{\beta}_{(M - 
 1)}(\a_s, \e)}~.
\label{idbet}
\eeq
Substituting \eq{idbet} into the definition of $R_{(M+1)}^{(M+1)}$,
and neglecting all terms ${\cal O} (\a_s^{M + 2})$ and higher, as well
as terms which are explicitly ${\cal O} (\e)$, one finds
\beq
 R_{(M+1)}^{(M+1)}(\a_s, \e) = R_{(M)}^{(M+1)} (\a_s, \e) +
 \left(\frac{\a_s}{\pi} \right)^{M + 1} \left( G_0^{(M + 1)} -
 \frac{b_{M - 1}}{4} G_1^{(1)} - \gtil^{(M + 1)} \right)~.
\label{diffr}
\eeq
The theorem is now proven if one can show that $R_{(M)}^{(M+1)}$ has
no poles in $\e$, since a constant at ${\cal O} (\a_s^{M + 1})$ can be
removed by suitably defining $\gtil^{(M + 1)}$. To see that this is
the case, note that $R_{(M)}^{(M+1)}(\a_s, \e)$ must, by the induction
hypothesis, be of the form
\beq
 R_{(M)}^{(M+1)}(\a_s, \e) = \sum_{m = 1}^M \sum_{l = 1}^\infty
 c_l^{(m)} \e^l \left(\frac{\a_s}{\p} \right)^m + \left(\frac{\a_s}{\p} 
 \right)^{M + 1} \sum_{p = p_0}^\infty d_p \e^p~.
\label{howr}
\eeq
Multiplying this times $\widehat{\beta}_{(M - 1)}(\a_s, \e)$, we must
get back the numerator of $R_{(M)}^{(M+1)}$, up to ${\cal O} (\a_s^{M
+ 2})$ corrections, namely
\beqa
 &&\left[1 + \frac{1}{4 \e} \sum_{n = 1}^{M - 1} b_{n - 1} 
 \left( \frac{\a_s}{\pi} \right)^n \right] \cdot 
 \left[ \sum_{m = 1}^M \sum_{l = 1}^\infty
 c_l^{(m)} \e^l \left(\frac{\a_s}{\p} \right)^m + \left(\frac{\a_s}{\p} 
 \right)^{M + 1} \sum_{p = p_0}^\infty d_p \e^p \right] = \nonumber \\
 && \hspace{5mm} = ~\sum_{n = 1}^M \sum_{m = 0}^\infty 
 G_m^{\left( n \right)} \e^m \left({{\a_s}\over{\p}}\right)^n - 
 \sum_{n = 1}^M \widetilde{G}^{(n)} \left({{\a_s}\over{\p}}\right)^n
 + {\cal O} (\a_s^{M + 2})~.
\label{leq}
\eeqa
Now, the right hand side has no poles in $\e$, and no term of order
$\a_s^{M + 1}$, thus on the left hand side all poles and all terms of
order $\a_s^{M + 1}$ must cancel. If $p_0 < 0$, on the other hand, one
will generate a term of the form $d_{-1} (\a_s/\p)^{(M + 1)} (1/\e)$
on the left hand side, which cannot be cancelled: in fact, note that
the first sum in $R_{(M)}^{(M+1)}$ starts at ${\cal O} (\e)$ by the
induction hypothesis, and $\widehat{\beta}_{M - 1}$ has only a simple
pole in $\e$. We conclude that $p_0 \geq 0$, as desired.

To find explicit expressions one has to implement the observation that
the coefficient of $\{ \left(\a_s/\p\right)^{M + 1}, \e^0 \}$ in
$R_{(M+1)}^{(M+1)}$ must vanish. One verifies that $\gtil^{(M + 1)}$
is given by the coefficient of $\{ \left(\a_s/\p\right)^{M + 1}, \e^0
\}$ in the ratio
\beq
 {\cal R}_{(M + 1)} (\a_s, \e) \equiv \frac{G_{(M + 1)} \left(\a_s, \e 
 \right)}{\widehat{\beta}_{(M)}(\a_s, \e)}~.
\label{whattoexpand}
\eeq
Explicitly,
\beq
 \gtil^{(M + 1)} = G_0^{(M + 1)} - {{b_0}\over{4}} 
 G_1^{(M)} - {{b_1}\over{4}} G_1^{(M - 1)}
 + {{b_0^2}\over{16}} G_2^{(M - 1)}
 - {{b_2}\over{4}} G_1^{(M - 2)}
 + {{b_0 b_1}\over{8}} G_2^{(M - 2)} + \ldots
\label{expli}
\eeq

\subsection{Real emission contributions}
\label{sec:rea-em}

The complete expression for the \msbns--scheme DY cross section in the
present framework is given by
\beq
 \widehat{\omega}_{\msb}(N) = \left|\frac{\Gamma(Q^2,
 \e)}{\Gamma(- Q^2, \e)} \right|^2 
 \left( \frac{\Gamma(-Q^2,\e)}{ \phi_V(\e)} \right)^2
 \left[U_R (N, \e) \left( \frac{\psi_R(N, \e)}{\phi_R(N, \e)} \right)^2 
 \right]~,
\label{comsb}
\eeq
where the factor in square brackets arises from real gluon emission.
Each function appearing in the real emission contribution
exponentiates: $\psi_R$ and $U_R$ according to Eqs.~(\ref{epsir}) and
(\ref{eur}), respectively, while $\phi_R$ is defined as the ratio of
\eq{phims} to \eq{phiv}. Renormalization group arguments can be
applied to each function, in $d = 4 - 2 \e$ dimensions, as described
in the Appendix. It is interesting to notice that running the coupling
in $d$ dimensions generates poles at two loops in the ratio $U_R
\psi_R/\phi_R$, although the input at one loop is finite.  The ratio
must however be finite to all orders, as a consequence of the
factorization theorem, together with the finiteness of the virtual
contributions demonstrated above. This poses constraints on the
two--loop coefficients of the functions involved, as described in more
detail in Section 4, tying together real and virtual contributions to
the cross section.

Collecting and organizing the exponential expressions of the real
emission functions, one can cast \eq{comsb} in the standard form
\beqa
 \widehat{\omega}_{\msb}(N) & = & \left|\frac{\Gamma(Q^2,
 \e)}{\Gamma(- Q^2, \e)} \right|^2
 \left( \frac{\Gamma(-Q^2,\e)}{ \phi_V(Q^2,\e)} \right)^2
 \exp \Big[ F_{\msbns} (\a_s) \Big]
 \nonumber \\ & \times & \exp \Bigg[ \int_0^1 \! dz \,
 \frac{z^{N - 1} - 1}{1 - z} \Bigg\{ 2 \,\int_{Q^2}^{(1 - z)^2 Q^2} 
 \frac{d \mu^2}{\mu^2} \, A \left(\a_s(\mu^2) \right) \nonumber \\
 & + & D \left(\a_s \left((1 - z)^2 Q^2 \right) \right) \Bigg\} 
 \Bigg] + {\cal O}(1/N)~.
\label{finms}
\eeqa
A one loop calculation, with the inclusion of running coupling
effects, yields
\beqa
 \label{msexp}
 && \hspace{-11pt}
 \log \left( \frac{\Gamma(-Q^2,\e)}{ \phi_V(Q^2,\e)} \right)
 = \frac{\a_s}{\pi} C_F \left(\frac{\zeta(2)}{4} - 2 \right) +
 \left( \frac{\a_s}{\pi} \right)^2 \left[ \frac{C_F b_0}{2} 
 \left(1 - \frac{3}{32} \zeta(2) - \frac{7}{24} \zeta(3) \right) + 
 \delta R^{(2)} \right]~,
 \nonumber \\
 && \hspace{-20pt}
 \quad F_{\msbns} \left( \a_s \right) = \frac{\a_s}{\pi} C_F
 \left(- \frac{3}{2} \zeta(2) \right) +
 \left( \frac{\a_s}{\pi} \right)^2 \left[
 - \frac{1}{4} b_0 C_F \left(1 - \frac{3}{8} \zeta(2) 
 - \frac{7}{4} \zeta(3) \right) + \delta F_{\msbns}^{(2)} \right]~,
 \eeqa
where $\delta R^{(2)}$ and $\delta F_{\msbns}^{(2)}$ are genuine
two--loop contributions, formally unrelated to running coupling
effects. Note that the function $D$ in \eq{finms} is the same as in
\eq{findis}: a non--trivial statement, due to the fact that such a
function, summarizing wide angle soft radiation, can be taken to
vanish in the threshold--resummed deep--inelastic structure function
\cite{Moch:2002sn,Forte:2002ni,Gardi:2002bk,Gardi:2002xm}.

We can now turn to a discussion of the relevance of the exponentiation
of constants in the two schemes, as tested at the two--loop level.

\section{Usage and limits of exponentiation}
\label{sec:two-loop-analysis}

It is clear that the results we have derived on the exponentiation of
$N$--independent terms do not have the predictive strength of the
standard resummation of threshold logarithms. In that case, in fact,
the pattern of exponentiation is highly nontrivial, and the
perturbative coefficients of entire classes of logarithms can be
exactly predicted to all perturbative orders performing just a low
order calculation. In the present case, even though constant terms
exponentiate, the determination of the exact value of the
$N$--independent contribution at, say, $g$ loops, always requires a
$g$--loop calculation, albeit in some cases a simplified one.

It remains true, however, that exponentiation and running coupling
effects generate contributions to all orders which originate from
low--order calculations. These contributions are an easily computable
subset of higher order corrections, and it is reasonable to use them
to estimate the full impact of higher orders. Specifically, given a
$g$--loop calculation of one of the cross sections we have discussed,
one can extract the value of the various functions appearing in the
exponent to that order, and then use exponentiation and RG running to
estimate the $(g+1)$--loop result. Given the existing results at two
loops~\cite{Hamberg:1991np,vanNeerven:1992gh}, one could construct an
estimate of the three--loop partonic cross section. Before embarking
in such a calculation, it is however advisable to test the case $g =
1$, {\it i.e.}  to use the two--loop results to verify the reliability
of the method, by comparing the exact results with the estimate
obtained by exponentiating the one--loop calculation and letting the
couplings run. To this end, we can expand the partonic cross section
in scheme $s$, $\widehat{\omega}_s (N)$ as
\beq
\widehat{\omega}_s (N) = \sum_{p = 0}^\infty \omega_s^{(p)} (N) 
\left(\frac{\a_s}{\pi} \right)^p~.
\label{perto}
\eeq
Next, we identify the coefficients of different powers of $\log N$,
by writing
\beq
\omega_s^{(p)} (N) = \sum_{i = 0}^{2 p} \omega_{s,i}^{(p)} 
\left(\log N + \gamma_E \right)^i~.
\label{pertp}
\eeq
So far we are dealing with the exact cross sections. Let now
$\widetilde{\omega}_{s,i}^{(p)}$ be the estimate for
$\omega_{s,i}^{(p)}$ obtained by evaluating the exponent exactly at
$p-1$ loops, adding running coupling effects, and expanding the result
to order $p$. One can define the deviation
\beq
\Delta \omega_{s,i}^{(p)} \equiv \frac{\omega_{s,i}^{(p)} -
\widetilde{\omega}_{s,i}^{(p)}}{\omega_{s,i}^{(p)}}~.
\label{dev}
\eeq
In computing our estimates for the DIS scheme, we will employ
\eq{findis}, with the one--loop results taken from \eq{coeff},
\eq{rattwo} and \eq{fdis}.  Furthermore, since we are taking into
account running coupling effects, we will include the terms
proportional to $b_0$ in Eqs.~(\ref{fdis2}) and~(\ref{myrat}). As far
as the \msbns scheme is concerned, we will make use of \eq{finms},
together with \eq{msexp}, with all purely two--loop contributions
defined to vanish. Exact results are taken from
Refs.~\cite{Hamberg:1991np,vanNeerven:1992gh}, focusing on ``soft and
virtual'' contributions (all other contributions are suppressed by
powers of $N$ at large $N$). To get numerical results we focus on
$SU(3)$ and set $n_f = 5$. The results for the deviations $\Delta
\omega_{s,i}^{(2)}$ in the two schemes are given in Table 1.
\TABLE{
\begin{tabular}{|c|c|c|c|c|c|} 
\hline
i & 0 & 1 & 2 & 3 & 4 \\ 
\hline
DIS & 0.26 & 1.17 & 0.13 & 0 & 0 \\ 
\hline
\msbns & \phantom{0}- 0.69\phantom{0}  & \phantom{0}1.79\phantom{0} & 
\phantom{0}0.33\phantom{0} & \phantom{.00}0\phantom{.00} & 
\phantom{.00}0\phantom{.00}\\
\hline
\end{tabular}
\caption{The deviations $\Delta \omega_{s,i}^{(2)}$, as defined in the
text, for the DIS and \msbns schemes.}
\label{tab1} }
To gain a little further insight, we also separate the contributions
proportional to the possible combinations of group invariants arising
at two loops ({\it i.e.} $C_F^2$, $C_A C_F$ and $n_f C_F$), and
display the results in Table 2 for the powers of $\log N$ which do not
lead to exact agreement ({\it i.e. $i = 0,1,2$}). Here we give the
coefficients separately for $\omega_{s,i}^{(2)}$ and
$\widetilde{\omega}_{s,i}^{(2)}$ for each scheme, since some of the
exact coefficients vanish, so that the corresponding deviations as
given in \eq{dev} are ill--defined.
\TABLE{
\begin{tabular}{|c|c|c|c|c|} 
\hline
& \multicolumn{2}{|c|}{\qquad \qquad \qquad DIS \qquad \qquad \qquad} &
  \multicolumn{2}{|c|}{\qquad \qquad \qquad \msb \qquad \qquad \qquad} \\ 
\hline
& \quad estimate \qquad & exact & \quad estimate \qquad & exact \\ 
\hline
\multicolumn{5}{|c|}{coefficients of $C_F^2$} \\ 
\hline
i = 0 & 38.06 & 39.03 & 3.33 & 1.79 \\
\hline
i = 1 & - 13.09 & - 14.41 & 0 & 0 \\
\hline
i = 2 & 9.85 & 9.85 & 5.16 & 5.16 \\
\hline
\multicolumn{5}{|c|}{coefficients of $C_A C_F$} \\ 
\hline
i = 0 & 5.63 & 18.12 & 13.12 & 6.82 \\
\hline
i = 1 & 9.83 & - 0.25 & 1.51 & - 0.47 \\
\hline
i = 2 & - 0.69 & 0.35 & 0 & 2.08 \\
\hline
\multicolumn{5}{|c|}{coefficients of $n_f C_F$}\\ 
\hline
i = 0 & - 1.02 & - 4.40 & - 2.38 & - 0.80 \\
\hline
i = 1 & - 1.79 & 0.35 & - 0.27 & - 0.52 \\
\hline
i = 2 & 0.12 & - 0.15 & 0 & - 0.56 \\
\hline
\end{tabular}
\caption{Comparison between estimates from exponentiation and exact
results at two loops, presented separately for different colour
structures, both in the DIS and \msbns schemes.}
\label{tab2} }

Several remarks are in order. From Table 1, we observe that, as
expected, leading ($i = 4$) and next--to--leading ($i = 3$) logs at
two loops are exactly predicted by one--loop results and running
coupling effects. Similarly, also as expected, NNL logarithms (in this
context $i = 2$) have a small discrepancy which is entirely traceable
to the two--loop cusp anomalous dimension $\gamma_K^{(2)}$. At the
level of $N$--independent terms ($i = 0$), the agreement is quite
reasonable, and in fact rather satisfactory in the DIS scheme, where
exponentiation accounts for three quarters of the exact answer.

Single log results, on the other hand, are much less satisfactory,
displaying a discrepancy larger than $100\%$ in both schemes.  The
reasons for this discrepancy are slightly different in the two
schemes, but they highlight the same generic problem. In the DIS
scheme, as discussed in the Appendix, running coupling effects in the
ratio of parton distributions $\psi/\chi$ generate a single log term
with the wrong sign with respect to the exact result. This term must
then be compensated by contributions which we would describe as
`genuine two--loop', which as a result display a rough proportionality
to $b_0$.

This phenomenon could be described as an `excess of factorization', in
the following sense: to achieve the accuracy and generality of
\eq{findis} it is necessary to introduce several functions, depending
on different scales. Not all of these dependencies, however, are
physical, and there may be (in fact there are) large cancellations in
the scale dependence between different functions. This fact has been
observed in the past: Catani and Trentadue pointed it out in
Ref.~\cite{Catani:1991rp}, and more recently a similar observation was
made by Gardi and Roberts in Ref.~\cite{Gardi:2002xm}. As a
consequence, approximate coefficients that are dominated by running
coupling effects, but not completely determined, may turn out to be
quite inaccurate.

In the \msb scheme, the same kind of cancellation is displayed in a
different way: there, as described in more detail below, one may use
the constraint imposed by the finiteness of the real emission
contributions to reexpress the single log coefficient in terms of
purely virtual functions, and in the process the weight of running
coupling effects changes considerably. Again, this indicates that
computed running coupling effects may easily be compensated by
unevaluated two--loop contributions.

Finally, one may observe in Table 2 that abelian ($C_F^2$)
contributions exponentiate with impressive accuracy, particularly in
the DIS scheme.  The slight superiority of the DIS scheme in this
regard appears to be a fairly generic feature in Tables 1 and 2,
perhaps to be ascribed to the more direct physical interpretation of
the subtractions, as compared to the \msb scheme.

To conclude, we would like to point out that the methods of
exponentiation we have outlined may be used not only for numerical
estimates, but also to obtain, or test, analytical
results. Specifically, since all functions employed have precise
diagrammatic definitions, the computation of certain coefficients at
two loops may be simplifed using this approach, as compared to a full
calculation of the cross section. Further, the factorization into
separately finite real and virtual contributions leads to constraints
connecting different coefficients, so that different two--loop results
can be nontrivially connected. To give an example, consider
\eq{comsb}.  There, the finiteness of the ratio $U_R \psi_R/\phi_R$ at
two loops imposes constraints tying together real and virtual
contributions to the cross section (recall that $\phi_R$ is defined as
$\phi_{\msbns}/\phi_V$).  Using the methods outlined in the Appendix
and imposing the cancellation of double poles in the ratio of real
functions, one may verify that the two--loop coefficient of $\log^2 N$
in $\widehat{\omega}_{\msbns} (N)$ must equal the two--loop cusp
anomalous dimension $\gamma_K^{(2)}$, as is well known. Further,
imposing the cancellation of single poles in the same ratio, one finds
that the value of the function $D$ at two loops is completely
determined by purely virtual diagrams. One finds
\beq
 \label{D2}
 D^{(2)} = \frac{3}{4} \zeta(2) b_0 C_F + 4 B_\delta^{(2)} 
 - 2 \gtil^{(2)}~,
\eeq
where $B_\delta^{(2)}$, the two--loop virtual part of the non--singlet
quark--quark splitting function~\cite{Herrod:1980rm,Curci:1980uw}, is
given by
\beq
 \label{B2}
 B_\delta^{(2)} = \frac{3}{2} C_F^2 \left(\frac{1}{16} - \frac{1}{2} 
 \zeta(2) + \zeta(3) \right) + \frac{C_A C_F}{4} \left(\frac{17}{24} + 
 \frac{11}{3} \zeta(2) - 3 \zeta(3) \right) 
 - \frac{n_f C_F}{6} \left(\frac{1}{8} + \zeta(2) \right)~,
 \eeq
while the second order contribution to the function $\gtil$ can be
determined via Eqs.~(\ref{expli}) and (\ref{mygam}), by matching the
resummed expression for $\Gamma(Q^2, \e)$ to the explicit results for
the dimensionally regularized one-- and two--loop Sudakov form
factors, as given for example in
Ref.~\cite{Matsuura:1989sm,Kramer:1987sg,Gonsalves:1983nq}. We find
\beqa
\label{varg}
 \gtil^{(2)} & = & G_0^{(2)} - \frac{b_0}{4} G_1^{(1)}~, \nonumber \\
 G_0^{(2)} & = & 3 \, C_F^2 \left(\frac{1}{16} - \frac{1}{2} \zeta(2) + 
 \zeta(3) \right) - \frac{C_A C_F}{4} \left(13 - \frac{11}{3} 
 \zeta(2) - \frac{2545}{108} \zeta(3) \right) \nonumber \\ && 
 - \, \frac{n_f C_F}{6} \left(\frac{209}{36} + \zeta(2) \right)~, \\
 G_1^{(1)} & = & C_F \left(4 - \frac{1}{2} \zeta(2) \right) \,. \nonumber
\eeqa
We have thus rederived the coefficient $D^{(2)}$, obtained earlier in
Refs.\cite{Contopanagos:1997nh,Vogt:2000ci} through matching to the
two--loop cross sections of
Ref.~\cite{Hamberg:1991np,vanNeerven:1992gh}, by using only
information from purely virtual contributions.

\section{Conclusions}
\label{conclu}

We have shown how to organize \textit{all} constants in the $N$--th
moment of the Drell-Yan cross section in the DIS and \msb schemes into
exponential forms. Our \msbns--scheme result has the special feature
that real and virtual contributions are separately finite. This
organization rests crucially upon the refactorization properties of
the unsubtracted Drell--Yan cross section and, for the DIS scheme, of
the non--singlet deep--inelastic structure function, near threshold
\cite{Sterman:1987aj}. For the \msb scheme the organization involves
the construction of an exponential series of pure counterterms that
cancels all divergences in the spacelike Sudakov form factor. We have
proven this cancellation to all orders. We emphasize that our
arguments imply exponentiation to the same degree of accuracy for the
\msbns--scheme DIS cross section $\widehat{F}_2 (N)$, although we have
not given a detailed evaluation in that case.

Although exponentiation of $N$--independent terms does not have the
same degree of predictive power as the resummation of threshold
logarithms, it can be used with some degree of confidence to gauge the
impact of higher order corrections to fixed order cross sections: we
found that $N$--independent contributions at two loops are reasonably
well approximated by the exponentiation of one--loop results. The
refactorization approach also leads to nontrivial connections between
real and virtual contributions to the cross section, which can be used
to test or in some cases simplify finite order calculations. On the
negative side, one cannot in general trust running coupling effects to
give by themselves a good approximation of two--loop results, unless
the various scales at which the couplings are evaluated are tied to
the physical scales of the full cross section.

It might be interesting to make full use of the available two--loop
information for the Drell--Yan and DIS cross sections to provide an
estimate of three--loop effects, along the lines of Section 4.  We
prefer to regard the techniques presented here, which extend threshold
resummations to a new class of terms, as a step towards the analysis
of yet other classes of perturbative corrections which might be
expected to exponentiate. A natural example is given by threshold
logarithms suppressed by an extra power of the Mellin variable $N$,
which have recently been analized in the case of the longitudinal DIS
structure function $F_L (N)$ in
Refs.~\cite{Akhoury:1998gs,Akhoury:2003fw}, and were shown to be
phenomenologically important in Ref.~\cite{Kramer:1996iq}.  Another
possible extension of our work is of course the study of
$N$--independent contributions to more complicated hard cross
sections, involving more colored particles, along the lines of
Refs.~\cite{Kidonakis:1997gm} and \cite{Sterman:2002qn}: this would be
an ingredient towards a precise resummed determination of such cross
sections, which would be of considerable phenomenological interest for
present and future hadron colliders.

\acknowledgments

The work of T.O.E. and E.L. is supported by the Foundation for
Fundamental Research of Matter (FOM) and the National Organization for
Scientific Research (NWO). L.M. was supported in part by the italian
Ministry of Education, University and Research (MIUR), under contract
2001023713--006.

%\bibliography{/user/t45/Projects/Bib/spires}
%\bibliographystyle{/user/t45/Projects/Bib/h-physrev3}

\vskip1cm

\appendix{\Large {\bf {\noindent Appendix}}}
\label{app}
\vskip 0.5cm
\renewcommand{\theequation}{A.\arabic{equation}}
\setcounter{equation}{0}

\noindent Let us briefly discuss the application of renormalization
group techniques to unconventional parton distributions such as the
functions $\psi(N, \e)$ and $\chi(N, \e)$ described in the text,
specifically focusing on the contributions involving real gluon
emission. The techniques of Ref.~\cite{Sterman:1987aj} lead to an
expression of the form
\beq
\label{a1}
 \psi_R(N, \e) = \exp \left\{ \int_0^1 dz ~\frac{z^{N - 1}}{1 - z} \int_z^1 
 \frac{dy}{1 - y} ~\kappa_\psi \left(\frac{(1 - y) Q}{\mu}, 
 \a_s(\mu^2), \e \right) \right\} + {\cal O} \left( 1/N \right)\,,
\eeq
and similarly for $\chi$. The functions $\psi_R$ and $\chi_R$ are both
renormalization group invariant ({\it i.e.} their respective anomalous
dimensions vanish) to this accuracy, for slightly different reasons:
$\psi_R$ cannot have overall UV divergences because its phase space is
restricted to fixed total energy emitted in the final state. This
automatically restrict also transverse momentum, so the phase space
integration is UV finite. $\chi_R$, on the other hand, has a phase
space restricted to fixed total light--cone momentum fraction, so that
in principle it may have UV divergences arising from transverse
momentum integrations. These divergences are in fact present, however
it can be shown that, at least at one loop and in the chosen axial
gauge, these divergences are suppressed by powers of $N$. Note that
this is not in contradiction with the fact that the divergent terms
for any quark distribution must be proportional to the
Altarelli--Parisi kernel. It simply means that the corresponding
divergences are of IR--collinear origin for the distributions at hand.

The consequence of this statement for the functions $\kappa_\psi$ and
$\kappa_\chi$ is that
\beq
\left(\mu \frac{\partial}{\partial \mu} + \beta \left(\e, 
\a_s \right) \frac{\partial}{\partial \a_s} \right) 
\kappa_\psi \left(\frac{(1 - y) Q}{\mu}, \a_s(\mu^2), \e 
\right) = 0~,
\label{renk}
\eeq
where $\beta \left(\e, \a_s \right)$ is the $\beta$ function in $d = 4
- 2 \e$. An identical equation is obeyed by $\kappa_\chi$.  Such
equations can be solved perturbatively to determine the dependence of
the distributions on the momentum scale. Consider again for example
$\kappa_\psi$, and let $\xi \equiv (1 - y) Q/\mu$. Expanding
\beq
\kappa_\psi \left( \xi, \a_s, \e \right) = \sum_{n = 1}^\infty
\left(\frac{\a_s}{\pi}\right)^n \kappa_\psi^{(n)} \left( \xi, \e 
\right)~,
\label{pexp}
\eeq
one easily finds that at the two loop level the coefficients must be
of the form
\beqa
\kappa_\psi^{(1)} \left( \xi, \e \right) & = & \kappa_{\psi, 0}^{(1)} 
\left(\e \right) \xi^{- 2 \e}~, \nonumber \\
\kappa_\psi^{(2)} \left( \xi, \e \right) & = & \kappa_{\psi, 0}^{(2)} 
\left(\e \right) \xi^{- 4 \e} + \frac{b_0}{4 \e}
\kappa_{\psi, 0}^{(1)} \left(\e \right) \xi^{- 2 \e}
\left(\xi^{- 2 \e} - 1 \right)~.
\label{expk}
\eeqa
explicit evaluation at one loop~\cite{Sterman:1987aj} yields
\beqa
\kappa_{\psi, 0}^{(1)} \left(\e \right) & = & 2 C_F \left(4 \pi 
\right)^\e \frac{\Gamma(2 - \e)}{\Gamma(2 - 2 \e)}~,
\nonumber \\
\kappa_{\chi, 0}^{(1)} \left(\e \right) & = & 2 C_F \left(4 \pi
\right)^\e \Gamma(2 + \e) \cos \left( \pi \e \right)~.
\label{onlog}
\eeqa
As observed in Section 2, the finiteness of the ratio $\psi_R/\chi_R$,
which is a consequence of factorization, requires that $\kappa_{\psi,
0}^{(1)} (\e) - \kappa_{\chi, 0}^{(1)} (\e) = {\cal O} (\e^2)$, since
the double integration in \eq{a1} generates a double pole. In fact,
upon redefing $\mu$ according to the \msbns prescription to absorb
factors of $\log (4 \pi)$ and $\gamma_E$,
\beq
\kappa_{\psi, 0}^{(1)} \left(\e \right) - 
\kappa_{\chi, 0}^{(1)} \left(\e \right) = 
2 C_F \e^2 \left[ 2 + \zeta(2) + \e \left( 4 + \zeta(2) -
2 \zeta(3) \right) + {\cal O} (\e^2) \right]~.
\label{oe1}
\eeq
One can go slightly further and observe that the finiteness of the
ratio $\psi_R/\chi_R$ also constrains the form of the pure two--loop
contribution to $\kappa_\psi$ and $\kappa_\chi$, given by the
functions $\kappa_{\psi, 0}^{(2)} (\e)$ and $\kappa_{\chi, 0}^{(2)}
(\e)$.  Specifically, inserting Eqs.~(\ref{pexp}) and (\ref{expk})
into \eq{a1}, and doing the same for $\chi_R$, one finds that the
ratio $\psi_R/\chi_R$ will develop a simple pole in $\e$ at two loops,
unless
\beq
\kappa_{\psi, 0}^{(2)} \left(\e \right) - 
\kappa_{\chi, 0}^{(2)} \left(\e \right) = 
\frac{3}{2} C_F b_0 \e \left( 2 + \zeta(2) \right) + 
\e^2 \delta \kappa_2^{(2)} + {\cal O} \left( 
\e^3 \right)~,
\label{oe2}
\eeq
in analogy with \eq{oe1}, with $\delta \kappa_2^{(2)}$ a constant
arising at two loops to be used below. This constraint also fixes the
coefficient of a contribution to the ratio proportional to $\log N$ at
two loops, {\it i.e.} at NNL level. To be precise one finds
\beqa
\left(\frac{\psi_R(N,\e)}{\chi_R(N,\e)}\right)^2 & = &
\exp \left[ \frac{\a_s}{\pi} C_F \left( 2 + \zeta(2) \right)
+ \left( \frac{\a_s}{\pi} \right)^2 \left( \frac{1}{8} \delta 
\kappa_2^{(2)} + \frac{1}{2} C_F b_0 \left( 2 + \zeta(2) \right) 
\left( \log N + \gamma_E \right) \right. \right. \nonumber \\
& - & \left. \left. \frac{3}{16} C_F b_0 \left( 4 + \zeta(2) 
- 2 \zeta(3) \right) \right) + {\cal O} \left( \e, \frac{1}{N}, 
\a_s^3 \right)\right]~.
\label{myrat}
\eeqa
Once again, the contributions arising at two loops should be taken
with a grain of salt when constructing the full cross section. It is
true in fact that in this way we have determined the leading
logarithmic contribution to this particular ratio, however in the full
cross section there are competing $\log N$ terms arising at two loops
from other functions, and in fact in the present case the logarithmic
term in \eq{myrat} goes in the wrong direction to `predict' NNL
logarithms at two loops, as discussed in the text. Similarly, there is
no guarantee that the uncalculated constant $\delta \kappa_2^{(2)}$
will not overwhelm the running coupling effects explicitly displayed
in \eq{myrat}.

\vspace{1cm}

%\bibliography{/user/t45/Projects/Bib/spires}
%\bibliographystyle{/user/t45/Projects/Bib/JHEP}

\providecommand{\href}[2]{#2}\begingroup\raggedright
\endgroup

\end{document}